\documentclass{sig-alternate}
\usepackage{hyperref}
\usepackage{enumitem}
\usepackage{subfigure}
\begin{document}
%
% --- Author Metadata here ---
\conferenceinfo{ASSETS}{'2013 Bellevue, Washington USA }
%\CopyrightYear{2013} % Allows default copyright year (20XX) to be over-ridden - IF NEED BE.
%\crdata{0-12345-67-8/90/01}  % Allows default copyright data (0-89791-88-6/97/05) to be over-ridden - IF NEED BE.
% --- End of Author Metadata ---

\title{Optimization of Switch Keyboards  }
%\subtitle{}

\numberofauthors{3} %  in this sample file, there are a *total*
\author{
% 1st. author
\alignauthor
Xiao (Cosmo) Zhang\\
       \affaddr{Department of Computer Science}\\
       \affaddr{Department of Psychological Sciences}\\
       \affaddr{Purdue University}\\
       \affaddr{703 Third Street, West Lafayette, IN 47907, USA}\\
       \email{zhang923@cs.purdue.edu}
% 2nd. author
\alignauthor
Kan Fang\\
       \affaddr{School of Industrial Engineering}\\
       \affaddr{Purdue University}\\
       \affaddr{315 N. Grant Street, West Lafayette, IN 47907, USA}\\
       \email{fang19@purdue.edu}
% 3rd. author
\alignauthor 
Gregory Francis\\
\affaddr{Department of Psychological Sciences}\\
       \affaddr{Purdue University}\\
       \affaddr{703 Third Street, West Lafayette, IN 47907, USA}\\
       \email{gfrancis@purdue.edu}
%\and  % use '\and' if you need 'another row' of author names
}
\date{20 June 2013}
% Just remember to make sure that the TOTAL number of authors
% is the number that will appear on the first page PLUS the
% number that will appear in the \additionalauthors section.
\maketitle
\begin{abstract}
Patients with motor control difficulties often ``type'' on a  computer using a switch keyboard to guide a scanning cursor to text elements. We show how to optimize some parts of the design of switch keyboards by casting the design problem as mixed integer programming. A new algorithm to find an optimized design solution is approximately 3600 times faster than a previous algorithm, which was also susceptible to finding a non-optimal solution. The optimization requires a model of the probability of an entry error, and we show how to build such a model from experimental data. Example optimized keyboards are demonstrated. 

\end{abstract}

%A category including the fourth, optional field follows...
\category{I.6}{SIMULATION AND MODELING}[Model Development]



\section{Introduction}
Some patients with spinal cord or brain injury lose motor control skills, even when they maintain their cognitive functions. In extreme cases of ``locked-in'' syndrome, patients are almost entirely paralyzed but remain conscious and need a means of communicating their thoughts. One method of communication is a switch keyboard with a scanning cursor that traverses a virtual keyboard. A binary switch  triggered by the patient (e.g., with an eye blink or a puff of air) guides the cursor to a character that is then typed on the computer.  Such typing is very slow (often measured in characters per minute rather than words per minute), and our goal is to make the process more efficient by identifying how to place characters on the keyboard in a way that allows for fast typing with few errors.\\
Francis and Johnson \cite{Francis2011} proposed that character placement could be treated as cost minimization that traded off speed and accuracy. Calculating these terms required a corpus of the kind of text that the patient would enter (e.g., poetry vs. HTML code), a model of errors, and an allowable average error rate. With such information, a hill-climbing algorithm could identify the cursor duration and key placements that minimized entry time.\\ 
\section{Mixed Integer Programming}
Here we show how to reframe the character placement problem as a mixed integer programming problem. A keyboard includes a set $\mathcal{I}$ of keys, and the cursor path scans across the rows, $\mathcal{J}$, and columns, $\mathcal{K}$, as guided by the user. Denote $f_{i}$ as the entry frequency of character $i$ in a given text of characters.  Set $x_{ijk}$ equal to $1$ if character $i$ is assigned to row $j$ and column $k$, and $0$ otherwise. Suppose the cursor moves across rows until the user triggers the switch to select a desired row. The cursor then moves across keys in the selected row until the user triggers the switch to select a particular key. Under such a cursor movement plan, the time to input a  character  at position $j,k$ is $t_{jk}=D(j+k)$, where $D$ is the duration of the cursor for each step. Assume $p_{jk}(D)$ is the proportion of errors when attempting to guide the cursor to position $j,k$, and that a user identifies an acceptable error rate of $\epsilon$.\\
Some characters are best assigned to fixed positions on the keyboard. For example,  the numbers of $0,1,\dots,9$ are traditionally grouped together in keyboard layout designs, so we can assign them as follows to the bottom of the keyboard: 
\begin{eqnarray*}
\mathcal{S} =\{(i,j,k)\vert (55, 7, 7),
(56, 7, 8), (57, 8, 1), (58, 8, 2), (59, 8, 3),\\
(60, 8, 4), (61, 8, 5), (62, 8, 6), (63, 8, 7), (64, 8,8)\}
\end{eqnarray*} 
where $i$ is the key index (in our example, keys 55--64 represent the numbers 0--9).  Then 
the problem of minimizing the entry time is as follows, given corpus size $n$, $D$, $\epsilon$, and $f_{i}$, 
\begin{equation}
\text{minimize}\ \ \mathbf{C}_{t}=\tfrac{1}{n}\sum\limits_{i\in \mathcal{I}}\sum\limits_{j\in \mathcal{J}}\sum\limits_{k\in \mathcal{K}}f_{i} t_{jk} x_{ijk} \label{d3} \tag{1a}
\end{equation}
subject to the following constraints:
\begin{align}
\mathbf{C}_{e}= \tfrac{1}{n}\sum\limits_{i\in \mathcal{I}}\sum\limits_{j\in \mathcal{J}}\sum\limits_{k\in
\mathcal{K}}f_i p_{jk}(D)x_{ijk} \le \epsilon \label{d4} \tag{1b}\\
\sum\limits_{j\in \mathcal{J}}\sum\limits_{k\in \mathcal{K}}x_{ijk}=1, \mbox{ for}\ i\in\mathcal{I};\label{d5}\tag{1c}\\
\sum\limits_{i\in \mathcal{I}}x_{ijk}=1, \mbox{ for}\ j\in\mathcal{J}; k\in\mathcal{K};\label{d6}\tag{1d}\\
x_s=1, \mbox{ for}\ s\in\mathcal{S}. \label{d7}\tag{1e}
\end{align}
Equations \eqref{d3} and \eqref{d4} are the definitions of the time cost and the error cost. Constraint \eqref{d4}
ensures that the error rate is not greater than the given acceptable threshold. Constraints
\eqref{d5} ensure that each key can be only assigned to one position. Constraints \eqref{d6} ensure that each position
can only contain one key. Constraints \eqref{d7} assign the number keys to their fixed positions.\\
The problem can be solved by using a Gurobi Optimizer with standard techniques that both guarantee a globally optimal solution (if it exists) and requires only about 3 seconds for a desktop computer. In contrast, the hill climbing algorithm on the same computer required approximately 3 hours to find a (not necessarily global) solution.\\
\section{Modeling error probabilities}
\begin{figure}[htb!]
		\centering
		\includegraphics[scale=0.25]{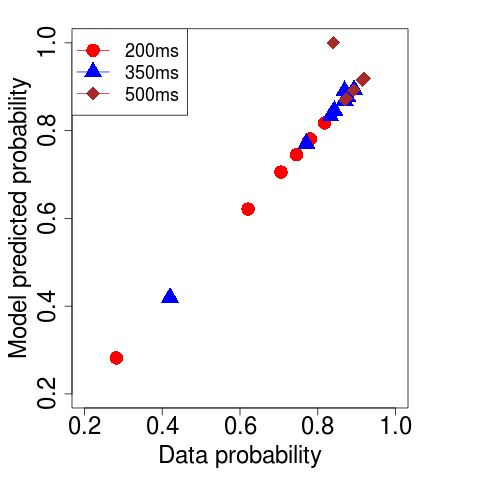}\\
		\caption{Scattergram of observed error probability and model fit}
		\label{cor}
\end{figure}

For the optimized keyboard to be useful, the $p_{jk}(D)$ terms need to accurately reflect the error probabilities of a user.  The task is to time a switch action to guide the cursor to a target key. We suspected that the timing of the switch action could be modeled with a Gamma distribution: $X \sim \Gamma(\kappa, \theta)$, where $\kappa$ is a shape parameter and $\theta$ is a scale parameter. The density function is denoted as $f(x; \kappa, \theta)=\frac{1}{\theta^{\kappa}}\frac{1}{\Gamma(x)}x^{\kappa-1}e^{-\frac{x}{\theta}},\ \text{for}\ x,\kappa,\theta>0$ where $\Gamma(x)$ is the gamma function evaluated at $x$, which is the elapsed time.\\
For each row in the keyboard and each cursor duration, we estimated the distribution parameters using data from Francis and Johnson \cite{Francis2011}. Their data indicated whether a switch response was generated early, correctly, or miss. A  Gamma distribution can produce similar probabilities by considering the area under the cdf prior to $x = Dj $, where $ j $ is the target row, between $Dj$ and $D(j+1)$, and after $D(j+1)$. We used a Nelder-Mead optimization method to estimate the distribution parameters. Figure~\ref{cor} shows a scattergram of the observed error proportions against the model fit. 

\begin{figure}[htb!]
\begin{minipage}[t]{0.5\linewidth}
%\centering
\includegraphics[scale=0.25]{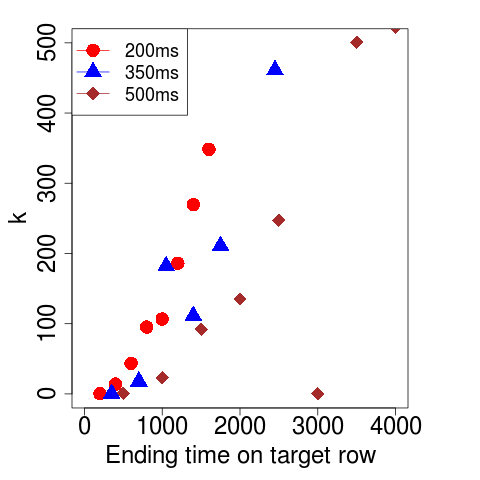}
\caption{$\kappa$-parameter}
\label{k-par}
\end{minipage}%
\begin{minipage}[t]{0.5\linewidth}
%\centering
\includegraphics[scale=0.25]{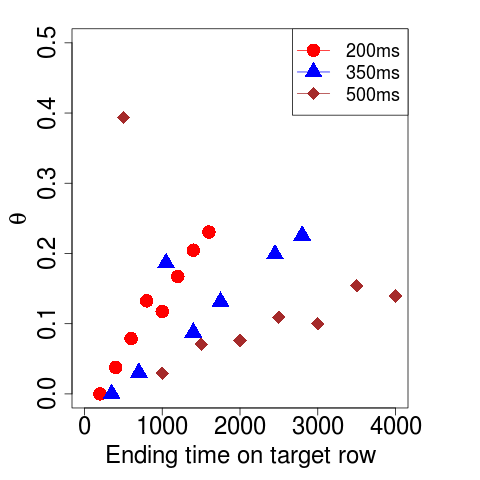}
\caption{$\theta$-parameter}
\label{theta-par}
\end{minipage}
\end{figure}

Figures~\ref{k-par} and \ref{theta-par} plot the $\kappa$ and $\theta$ estimates as a function of elapsed time and cursor duration. They suggest systematic patterns that will be modelled in future work in the Gamma distributions. 

%\begin{figure}
%  \centering
%  \subfigure[k-parameter]{
%    \includegraphics[scale=0.25]{k_par.png}}	
%  \hspace{0.01cm}
%  \subfigure[$\theta$-parameter]{
%    \includegraphics[scale=0.25]{theta_par.png}}
%    \caption{Parameters of the Gamma Distribution Modelling}
%	\label{par}	
%\end{figure}

\section{Optimization Results}
\begin{figure}[htb!]
  \centering
  \subfigure[$\epsilon = 0.30$]{
    \includegraphics[scale=0.2]{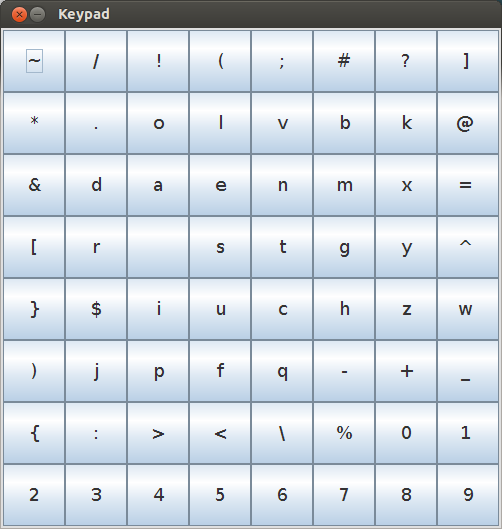}}	
  \hspace{0.05cm}
  \subfigure[$\epsilon = 0.50$]{
    \includegraphics[scale=0.2]{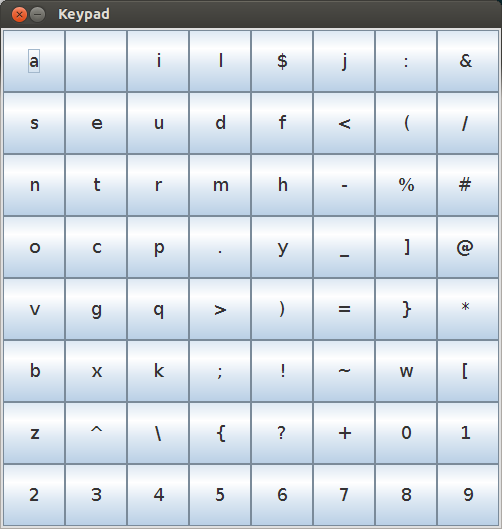}}
    \caption{Two optimized keyboards}
	\label{optr}	
\end{figure}

Figure~\ref{optr} shows two optimal keyboards that trade off speed and accuracy for a fixed cursor duration $D=0.35$ seconds. The keyboard on the left used $\epsilon=0.30$ and so emphasizes character placement to minimize the error rate when entering a given text corpus (a set of famous quotes). The keyboard on the right used $\epsilon=0.50$ and so emphasizes character placement to minimize the time of entry.

\section{Conclusions}
We identified how to solve the problem of optimal character placement on a switch keyboard using mixed integer programming. The algorithm is orders of magnitude faster than a previous approach and will allow for consideration of many other keyboard properties (e.g., various cursor paths, different switch devices, shortcuts). The algorithm requires an accurate model of error rates, which must be based on human data. We described a model that matches the observed data and has promise for future investigations. 
%\end{document}  % This is where a 'short' article might terminate

%ACKNOWLEDGMENTS are optional
\section{Acknowledgments}
This work was supported by a Project Development Team within the ICTSI NIH/NCRR Grant Number RR025761.

\bibliographystyle{abbrv}
\bibliography{all}
% sigproc.bib is the name of the Bibliography in this case

\end{document}